\documentclass[prb,twocolumn]{revtex4}   

\usepackage{graphicx}

\begin{document}

\title
{
Quantum phase transitions in the extended periodic Anderson model}

\author
{
Akihisa Koga,$^{1,2}$ Norio Kawakami,$^{1}$ 
Robert Peters$^2$ and Thomas Pruschke$^2$
}

\affiliation
{
$\it ^1$Department of Physics, Kyoto University, Kyoto 606-8502, 
Japan\\
$\it ^2$Institut f\"ur Theoretische Physik Universit\"at G\"ottingen,
G\"ottingen D-37077, Germany
}

\date{\today}

\begin{abstract}
We investigate quantum phase transitions in the extended periodic Anderson
model, which includes electron correlations within and between itinerant 
and localized bands. We calculate zero and finite temperature 
properties of the system using
the combination of dynamical mean-field theory and
the numerical renormalization group.
At half filling, 
a phase transition between a Mott insulating state 
and a Kondo insulating state occurs in the strong coupling regime.
We furthermore find that 
a metallic state is stabilized in the weak coupling regime.
This state should be adiabatically
connected to the orbital selective Mott state with one orbital
localized and the other itinerant. 
The effect of hole doping is also addressed.
\end{abstract}

\pacs{Valid PACS appear here}%

\maketitle

\section{Introduction}
Strongly correlated electron systems with orbital degeneracy 
have attracted considerable interest. 
One typical example is the transition metal oxide $\rm LiV_2O_4$.
In this compound, unexpected heavy fermion-like behavior was observed,
\cite{Kondo} 
which is quite likely induced by an interplay between 
the geometrical frustration of the spinel structure 
and the orbital degeneracy 
of the $d$ electrons.\cite{Kaps,Isoda,Fujimoto,Yamashita,Tsunetsugu}
One particular scenario suggests that due to the V positions 
in the local trigonal crystal environment,
the $t_{2g}$ states at the Fermi energy are split into $a_{1g}$ and $e_g^\pi$
subbands, which then play different roles in 
stabilizing the heavy electron state.\cite{AnisimovLiVO,Kusunose}
For example, due to different bandwidths of the 
$a_{1g}$ 
and $e_{g}^\pi$ bands, 
the local Coulomb correlations can induce a Mott insulating
state in the former while the latter stay metallic.\cite{AnisimovLiVO,Arita}
A hybridization between the two subshells on neighboring $V$ atoms in the
unit cell then
can explain the observed heavy fermion behavior.
Other examples are the compounds
$\rm Ca_{2-x}Sr_{x}RuO_4$\cite{Nakatsuji} and 
$\rm La_{n+1}Ni_nO_{3n+1}$,\cite{Sreedhar,Kobayashi}
where a similar orbital-selective Mott transition\cite{Anisimov} 
is suggested to be induced by
the chemical substitutions and the change of temperature, respectively.
It was furthermore reported that a heavy fermion state is realized 
in the former compound around $x\approx 0.2$,\cite{Nakatsuji} while
a bad metal
with localized spins at low temperatures is characteristic 
of the latter.\cite{Kobayashi}
These observations stimulate further theoretical investigations 
on multiorbital systems\cite{KogaV,Medici,Liebsch,KogaOSMT,Inaba,OSMT}

A common feature of those interesting examples is that
localized and itinerant electrons and their correlations 
in the multiorbital system seem to play an important role
in the formation of heavy fermion states in transition metal oxides.
Motivated by this, we here investigate a multiorbital system 
with localized and itinerant bands.
It is known that 
the hybridization together with local electron correlations 
leads to the Kondo effect and in turn to the large density of states 
at the Fermi level, which
explains the heavy fermion states observed in rare-earth
compounds.\cite{Grewe91,Yamada,Hewson} 
On the other hand, electron correlations in an itinerant band also yield 
a heavy mass and eventually induce a transition
to the Mott insulating state.\cite{Brinkman,Georges}
Thus the interesting question arises 
whether the heavy fermion states originating from the Kondo effect and close
to the Mott transition can be distinguished from each other or not.
This question has recently been discussed by several authors,%
\cite{KogaV,Medici,Sato,Schork}
but it was not clear 
how the heavy fermion state is realized at low temperatures.
In particular, the role of the Hund coupling
needs to be clarified, which may be important 
in $f$-electron systems as well as transition metal oxides.
Therefore,
it is necessary to discuss low temperature properties 
in correlated multiorbital electron systems 
with localized and itinerant bands systematically.

To this end, we investigate an extended version of the periodic Anderson model,
where not only onsite Coulomb interactions in the localized and itinerant bands
but also the interband Coulomb interaction and Hund coupling are 
taken into account. 
The properties we are interested in -- heavy fermion behavior and Mott
transition -- are well captured by the 
dynamical mean field theory (DMFT),\cite{Metzner,Muller,Georges,Pruschke} 
and furthermore do not depend critically on details of the band structure.
Thus we will discuss the low-temperature properties for a
particle-hole symmetric system within DMFT. 
We will clarify that the Hund coupling between different orbitals 
plays an important role, leading to a quantum phase transition 
between a Kondo insulating phase, a Mott insulating phase and 
the metallic phase. 
The Mott and Kondo insulators also behave differently 
when doping the system. 

The paper is organized as follows. 
In Sec. \ref{sec2}, we introduce the extended periodic Anderson model and
briefly summarize DMFT and numerical techniques.
The competition between some phases is discussed in Sec. \ref{sec3}.
We also discuss the effect of hole doping in Sec. \ref{sec4}. 
A brief summary is given in Sec. \ref{sec5}.

\section{Model and Method}\label{sec2}

\begin{figure}[htb]
\begin{center}
\includegraphics[width=6cm]{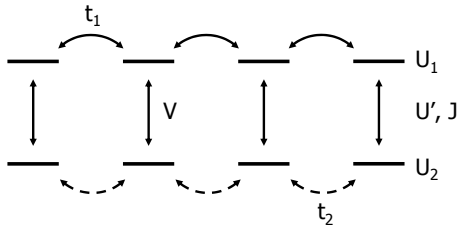}
\end{center}
\vskip -4mm
\caption{Sketch of the structure of the model Hamiltonian}
\label{fig:model}
\end{figure}

We consider a correlated electron system with orbital degeneracy
described by the Hamiltonian
\begin{widetext}
\begin{eqnarray}
{\cal H}&=&H_t + \sum_i H_{local}^{(i)},\\
H_t&=&\sum_{\langle ij\rangle \alpha\sigma}
\left[ t_{ij}^{(\alpha)}-\mu \delta_{ij}\right]
c_{i\alpha\sigma}^\dag c_{j\alpha\sigma},\\
H_{local}^{(i)}&=&V\sum_{\sigma}\left( 
c_{i1\sigma}^\dag c_{i2\sigma}+c_{i2\sigma}^\dag c_{i1\sigma}
\right)
+\sum_{\alpha} U_\alpha n_{i\alpha\uparrow} n_{i\alpha\downarrow}+
\sum_{\sigma\sigma'}\left(U'-J\delta_{\sigma\sigma'}\right) 
n_{i1\sigma}n_{i2\sigma'}\nonumber\\
&-&J\left[ 
\left( c_{i1\uparrow}^\dag c_{i1\downarrow} 
c_{i2\downarrow}^\dag c_{i2\uparrow}
+ c_{i1\uparrow}^\dag c_{i1\downarrow}^\dag 
c_{i2\uparrow} c_{i2\downarrow}\right)+\mbox{h.c.} \right]\label{eq:model}
\end{eqnarray}
\end{widetext}
where $c_{i\alpha\sigma}^\dag (c_{i\alpha\sigma})$ 
creates (annihilates) an electron 
with spin $\sigma(=\uparrow, \downarrow)$ and orbital
index $\alpha(=1, 2)$ at the $i$th site, 
and $n_{i\alpha\sigma}=c_{i\alpha\sigma}^\dag c_{i\alpha\sigma}$.
For orbital $\alpha$,
$t_{ij}^{(\alpha)}$ represents the transfer integral,
$V$ the hybridization between orbitals,
$U_\alpha$ and $U'$ the intra-orbital and inter-orbital Coulomb interactions,
$J$ the Hund coupling, and 
$\mu$ the chemical potential.
The structure of the model Hamiltonian is schematically shown in Fig. \ref{fig:model}.
We note that the model is reduced to conventional systems
in several limiting cases. 
For example, when $V=0$,
the system becomes the two orbital Hubbard model,
where we expect a Mott transition to occur.
On the other hand, when $t_{ij}^{(2)}=0$, we recover at $U_1=U'=J=0$
the conventional periodic Anderson model for 
heavy-fermion systems,\cite{Hewson} 
while the choice $U_1=U_2=V=0$, $J>0$ leads to the double exchange model
discussed in connection with magnetism in transition metal oxides.
\cite{Zener,Anderson,Kubo,Furukawa}

In this paper, we want to focus on the low-temperature properties of a system
with one band localized and one kept itinerant, {\it i.e.}\
we choose the hopping integral for the $\alpha=2$ band as $t_{ij}^{(2)}=0$, but
allow all other parameters to be finite.
To treat the extended periodic Anderson model in this parameter regime,
we use the DMFT. This method has been developed in several groups
\cite{Metzner,Muller,Georges,Pruschke}
and has successfully been applied to correlated electron systems such as
the single-band Hubbard model,\cite{Caffarel,OSakai,single1,single2,single3,single4,BullaNRG} 
the two-band Hubbard model\cite{2band1,2band2,Koga,Momoi,OnoED,multi,Ono4,Pruschke2,KogaOSMT,OSMT}
or the periodic Anderson model.\cite{PAM,Mutou,Saso,pam_pr,Sato,Ohashi,Medici}
In the DMFT, a lattice model is mapped to an effective impurity model,
where local electron correlations are taken into account precisely.
The requirement that the site-diagonal lattice Green function is equal
to the one of the effective quantum impurity then leads to a self-consistency
condition for the parameters entering the impurity problem.
This treatment is formally exact for infinite spatial dimensions. If one is
allowed to ignore nonlocal correlations, {\it e.g.}\ sufficiently far away from
phase transitions with long-range order, the method can be used as a quite
reliable and accurate approximation to three dimensional systems.

Within the DMFT, the lattice typically enters only via the density of states
(DOS) of the system with vanishing two-particle interactions. Since we
are interested in generic features of the extended periodic Anderson model,
we are free to choose a convenient lattice structure. Here, we use
a Bethe lattice with infinite coordination,\cite{Georges} for which the
DMFT self-consistency equation
is simplified to\cite{Sato,Schork}
\begin{eqnarray}
\left[ \hat{G}_0^{-1}(z)\right]_{11} &=& z+\mu-\left( \frac{D}{2} \right)^2 
\left[ \hat{G}_{loc}(z)\right]_{11},
\label{eq:condition}
\end{eqnarray}
where $\hat{G}_0$ is the non-interacting Green function 
for the effective impurity
model and $\hat{G}_{loc}$ is the local Green function. Both are matrices
with respect to the orbital index, but due to $t_{ij}^{(2)}=0$ only the
$(1,1)$ element is needed. As is well-known,\cite{Georges}
the relation (\ref{eq:condition}) is equivalent to a 
semi-elliptic DOS
for the itinerant band $(\alpha=1)$ with half-bandwidth $D$.

Let us note that the
localized band $(\alpha=2)$ does not appear explicitly in the self-consistency
condition eq.\ (\ref{eq:condition}).
For the effective impurity model entering the DMFT, this means that
only one of the impurity orbitals
connects to the rest of the system through an effective hybridization function
defined by
$\left(\frac{D}{2}\right)^2\left[\hat{G}_{loc}(z)\right]_{11}$. 
In the following, this quantity is represented 
by an auxiliary set of dynamical
degrees of freedom, which we will call an effective bath.

The Hamiltonian representing the effective impurity model can now be
written as
\begin{eqnarray}
H&=&\sum_k \epsilon_k 
a_{k\sigma}^\dag a_{k\sigma}
+\sum_{k\sigma}\gamma_k\left( 
a_{k\sigma}^\dag c_{1\sigma}+c_{1\sigma}^\dag a_{k\sigma}
\right)\nonumber\\
&+&\sum_{\alpha\sigma} E_{\alpha} n_{\alpha\sigma}
+H_{local}
\;\;,\label{eq:imp}
\end{eqnarray}
where $a_{k\sigma}$ are the auxiliary annihilation operators 
with quantum number $k$ and spin $\sigma$ defining 
the effective bath,
$E_{\alpha}=-\mu$ 
is the energy level for the impurity site,
$n_{\alpha\sigma}=c_{\alpha\sigma}^\dag c_{\alpha\sigma}$
and $H_{local}$ the local interaction term as defined in (\ref{eq:model}).
The set of parameters for the effective bath $\{\epsilon_k, \gamma_k\}$ 
must be determined self-consistently through the DMFT condition
eq.\ (\ref{eq:condition}). 

The Hamiltonian (\ref{eq:imp}) represents a quantum impurity model, which
is a challenging theoretical problem on its own account. Thus,
to discuss low-temperature properties of the extended periodic
Anderson model in the DMFT, we need a tool to accurately solve such
quantum impurity models. While applicable in the weak-coupling regime,
perturbation theory in general fails 
in the strong coupling regime, where for example the anticipated
transition between Mott and Kondo physics should appear. 
Exact diagonalizations and quantum Monte Carlo (QMC) simulations
are known to be numerically exact methods,
but both cannot properly resolve exponentially small energy scales, which
again will play an important role in the interesting regime, where
Kondo and Mott physics compete.
In addition, due to the required computational resources, 
QMC simulations typically can access only a 
restricted window of the parameter space and 
can also be subject to a severe sign problem when applied
to complex multi-orbital models.

Therefore, we here use Wilson's numerical renormalization group\cite{nrg_rmp}
(NRG) to solve the effective impurity model.
In the NRG, one discretizes the effective bath on a logarithmic mesh
by introducing a discretization parameter $\Lambda>1$. The resulting discrete
system can be mapped to a semi-infinite chain with exponentially decreasing
couplings,\cite{nrg_rmp} which  
allow us to access and discuss properties involving exponentially small energy
scales
quantitatively. To ensure that sum rules for dynamical quantities are
fulfilled,
we use the complete-basis-set algorithm proposed recently.\cite{anders,Peters,weichselbaum:06}
We observe that obeying the sum rules is mandatory to properly describe the
low energy properties of the system away from half filling.
In the NRG calculations, we use a discretization parameter $\Lambda=2$ and
keep $1200$ states at each step.

\section{Results at half filling}\label{sec3}
In the following, we will present results calculated for
$U=U_1=U_2$ with $J/U=0.1$ and $U'=U-2J$ fixed. 
The effect of Hund's coupling will be addressed later.

Let us begin with the low-temperature properties 
of the extended periodic Anderson model at half filling, which is realized
by fixing the chemical potential $\mu=\mu_0=U/2+U'-J/2$.
For the conventional periodic Anderson model $U_1=U'=J=0$, 
it is known that the band insulator at $(V/D\gg 1)$ 
is adiabatically connected to the Kondo insulator with
$U_2/D\gg 1$.
Furthermore, it was claimed that the introduction of 
a Coulomb interaction for the conduction band $U_1$ 
does not induce a transition to the Mott insulating state.\cite{Sato,Schork}
Namely, in this case always antiferromagnetic spin correlations between the
orbitals develop. However, in the presence of
Hund's coupling $J$, which induces
{\it ferromagnetic} spin correlations between the orbitals,
one should expect a competition between the singlet formation due to the
hybridization 
and a high spin formation due to the Hund coupling.

\begin{figure}[htb]
\begin{center}
\includegraphics[width=7cm]{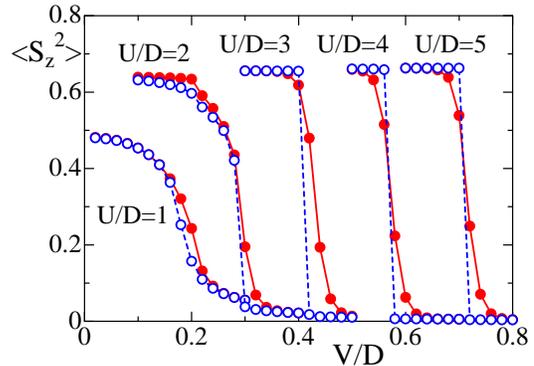}
\end{center}
\vskip -4mm
\caption{(color online) The local squared moment $\langle S_z^2 \rangle$ as a function of 
the hybridization $V/D$. Solid (open) circles represent the results at  temperature $T/D=2.1\times 10^{-2}$ $(6.5\times 10^{-4})$. }
\label{fig:sz}
\end{figure}

To discuss how spin correlations develop between the degenerate orbitals,
we calculate the squared spin moment $\langle S_z^2 \rangle$, 
where $S_z=\sum_\alpha \left( n_{\alpha\uparrow}-n_{\alpha\downarrow} \right)/2$,
as shown in Fig. \ref{fig:sz}. 
When $V/D$ is large, the local spin moment is strongly suppressed,
$\langle S_z^2\rangle \to 0$. In this regime, 
the hybridization together with the intraorbital Coulomb interactions
leads to the formation of an interorbital singlet, 
where the insulator (Kondo insulator) will be realized 
in the half-filled particle-hole symmetric system.
On the other hand, different behavior occurs for small $V/D$,
as shown in Fig. \ref{fig:sz}.
Here we find that $\langle S_z^2 \rangle \to 2/3$ when $U/D\gtrsim2$,
while $\langle S_z^2 \rangle \to 1/2$ in the other limit.

In the former case, strong correlations localize the electrons in 
band $\alpha=1$. Note
that due to a finite Hund's coupling, the corresponding Mott transition
will occur at a smaller value of $U$ as compared to the one-band Hubbard model
here.\cite{OnoED,Pruschke2}
Furthermore, for small $V$,
the Hund's coupling will win over the effective antiferromagnetic exchange
$\sim V^2/U$
generated by the hybridization and the Coulomb interaction, 
{\it i.e.}\ a spin triplet state $S=1$ is realized, 
which implies $\langle S_z^2\rangle=2/3$.
Since for large $V/D$ the ground state is a local singlet with $S=0$ , we expect
that at $T=0$ a phase transition
from the Mott insulating state to the Kondo insulating state
occurs as a function of $V/D$.
In fact, the phase transition is clearly visible for the lower temperature,
as shown in Fig.\ \ref{fig:sz}.

On the other hand, in the regime $U/D\lesssim 2$, we observe a smooth curve 
for $\langle S_z^2\rangle$ as a function of $V$, which
furthermore depends only little on temperature,
except for a small region around the turning point. This behavior is
shown in Fig.\ \ref{fig:sz} for $U/D=1$ as example.
For small $V/D$ Hund's exchange will again dominate over the antiferromagnetic
interorbital coupling generated by $V$. Thus we are effectively left with a
Fermi liquid ferromagnetically coupled to a localized spin, {\it i.e.}\ a 
ferromagnetic Kondo model. As is well-known, this model has a metallic ground
state,\cite{Hewson} where the local spin acts as a potential scatterer.
The increase of $V$ eventually leads to a dominance of the hybridization term 
and
a Kondo screening of the local spin by the itinerant electrons. For finite
temperatures one will observe a crossover, because only for $T_{\rm K}\gg T$
a full singlet with $\langle S_z^2\rangle=0$ between local and band spins is
realized. 
With decreasing temperature, the crossover becomes sharper and, because
one now probes lower Kondo temperatures,
shifts to the left, as seen in Fig.~\ref{fig:sz}. Eventually it turns into
a transition at $T=0$, where the Kondo state always leads to a singlet for
arbitrary small antiferromagnetic coupling.

\begin{figure}[htb]
\begin{center}
\includegraphics[width=8.4cm,clip]{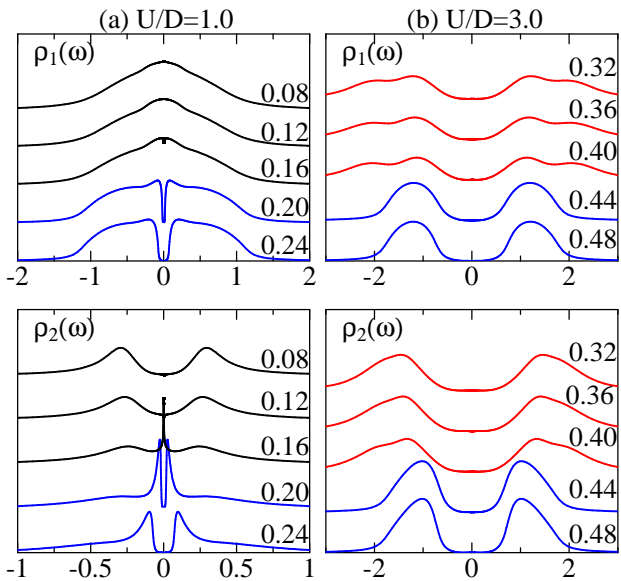}
\end{center}
\vskip -4mm
\caption{(color online) Density of states for 
the extended periodic Anderson model 
with fixed $U/D=1.0$ (a) and $U/D=3.0$ (b) at $T/D=6.5\times 10^{-4}$. 
The numbers represent the values $V/D$.
}
\label{fig:dos-12}
\end{figure}
\begin{figure}[htb]
\begin{center}
\includegraphics[width=7cm,clip]{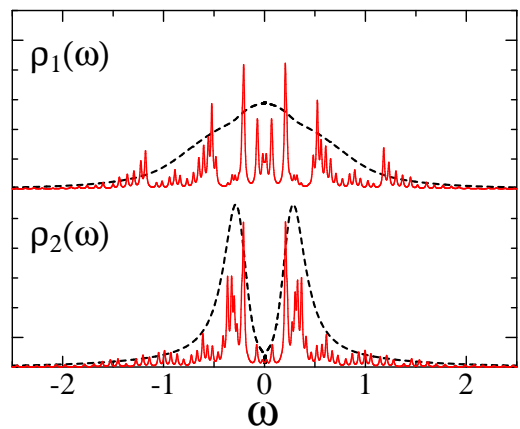}
\end{center}
\vskip -4mm
\caption{(color online)
Solid and dashed lines represent the density of states 
obtained by exact diagonalization with 12 sites
and the NRG method at $T/D\approx 10^{-8}$
for the metallic state with $U/D=1.0$ and $V/D=0.1$.
}
\label{fig:dos}
\end{figure}

This qualitative discussion can be further substantiated by looking at 
the density of states for itinerant and localized orbitals 
in Fig.\ \ref{fig:dos-12}.
We first focus on the case $U/D=1$. 
As anticipated before, for small $V/D$ a metallic state is realized 
in the orbital $\alpha=1$. The orbital $\alpha=2$, on the
other hand remains localized.
To confirm the stability of the metallic state and support the
qualitative discussion,
we also show the density of states at a much lower temperature 
$T/D\approx10^{-8}$
as the dashed lines in Fig. \ref{fig:dos}.
As expected, we find that the metallic state is certainly 
the ground state for the $\alpha=1$ orbital,
while a strongly suppressed DOS
at Fermi level is seen 
in the other orbital.
The NRG results are consistent with those obtained from
a calculation using exact diagonalization,\cite{Georges,Caffarel} 
see full lines in Fig.\ \ref{fig:dos}.
Further increase of the hybridization beyond the phase boundary 
$(V/D)_c\approx 0.14$ eventually leads to the appearance
of a Kondo resonance in $\rho_2(\omega)$, as shown in Fig.\ \ref{fig:dos-12} (a),
and a corresponding
reduction of the DOS at the Fermi energy in $\rho_1(\omega)$. 
For even larger $V$, one then finds the structures
known from Kondo insulators,\cite{PAM,grewe_ssc,pam_pr} {\it i.e.}\ a Kondo
resonance split by a hybridization gap in $\rho_2(\omega)$
and a gap of the same width in $\rho_1(\omega)$.
Note that one would actually expect a similar gap to appear in $\rho_2(\omega)$
at $V/D=0.16$. However, this gap appears only for $T\ll T_{\rm K}$ and is
thus not yet visible here.

When $U/D=3$, on the other hand,
we always find an insulating state, which can be characterized as a Mott 
(Kondo) insulator with a charge gap in both orbitals
for small (large) $V/D$.
When $V$ approaches the critical value $(V/D)_c \approx 0.42$ from below,
the charge gap around the Fermi level decreases,
but does not seem to vanish even at the critical point,
which implies that the phase transition is of first order, in accordance with
the behavior of $\langle S_z^2\rangle$. 
The initial reduction of the charge gap with
increasing $V$ can be interpreted in the following way. 
For small $V$, the
Mott state is stabilized at a lower critical $U_c$ due to ferromagnetic Hund's
coupling.\cite{OnoED,Pruschke2} With increasing $V$, the additionally generated
antiferromagnetic exchange effectively reduces this effect, 
{\it i.e.}\ the orbital
$\alpha=1$ is driven closer to its critical value and the charge gap decreases.
As soon as the antiferromagnetic coupling dominates, the gap scale is set
by the Kondo scale. The properties very close and at the critical point are
thus determined by this subtle competition of energy scales. Whether the
two regimes are separated by a quantum critical point or the transition is
driven by a simple level crossing remains to be investigated.

\begin{figure}[htb]
\begin{center}
\includegraphics[width=7cm,clip]{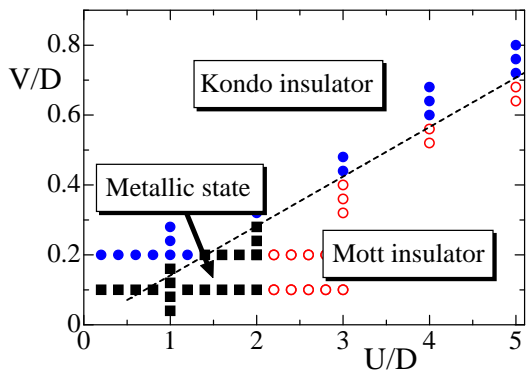}
\end{center}
\vskip -4mm
\caption{(color online) Finite temperature phase diagram for 
the extended periodic Anderson model with
$J/U=0.1$ and $T/D=6.5\times 10^{-4}$. 
Solid circles, open circles and solid squares denote
the Kondo insulator, the Mott insulator and the metallic state, respectively. 
The dashed line is obtained from the simplified model
(see text).
}
\label{fig:phase30}
\end{figure}

By performing similar calculations for various parameter values, 
we obtain the phase diagram shown in Fig. \ref{fig:phase30}.
When $V/D$ is large, the Kondo insulator prevails,
where the spins for the localized band $(\alpha=2)$ are screened
by forming a singlet with the electrons in orbital $\alpha=1$.
The increase of the Coulomb interaction $U$ with fixed $J/U$ 
decreases the hybridization gap characteristic of the Kondo insulating
state. Finally, a phase transition occurs to 
the Mott insulating state with a local moment $S=1$. Note that in this regime
the Kondo interaction is dominated by ferromagnetic Hund's exchange. 
The competition between these phases can be qualitatively described 
by a simplified local Hamiltonian for two orbitals,
since the bandwidth has little effect on the phase transition.
By examining the lowest state for this simplified model,
we find a level crossing at
$V/D=\sqrt{2}J/D$ between the singlet and the triplet states,
which is shown as the dashed line in Fig. \ref{fig:phase30}.
Obviously, this line is in good agreement with 
the phase boundary obtained by DMFT. This result also suggests that
the actual phase transition is of first order and not accompanied by
a quantum critical point.
 
\begin{figure}[htb]
\begin{center}
\includegraphics[width=8cm]{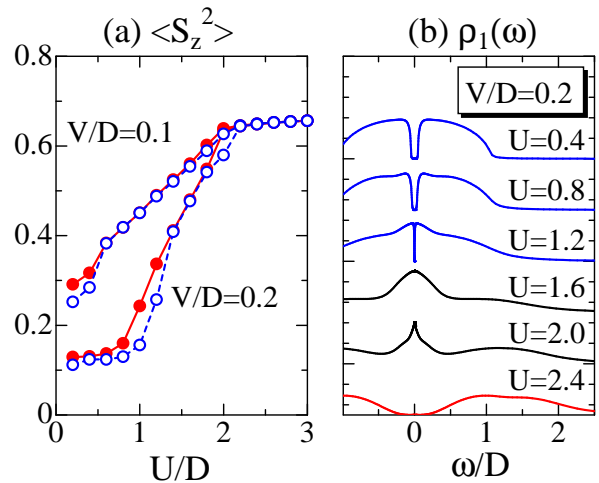}
\end{center}
\vskip -4mm
\caption{(color online)
(a) Local moment as a function of the Coulomb interaction $U$
for $V/D=0.1$ and $0.2$. 
Solid (open) circles are the results for $T/D=2.1\times 10^{-2} 
(6.5\times 10^{-4})$.
(b) Density of states for the orbital $\alpha=1$ and different values
$U/D=0.4$, $0.8$, $1.2$, $1.6$, $2.0$ and $2.4$ 
for $T/D=6.5\times 10^{-4}$.
}
\label{fig:Mott}
\end{figure}

When the hybridization is small $(V/D<\sqrt{2}JD)$ and $U/D\lesssim 2$, 
a metallic state appears with
orbital $\alpha=2$ almost localized and $\alpha=1$ itinerant. 
As the Coulomb interaction is increased for the metallic state,
the quasi-particle peak becomes sharper
for the itinerant orbital, as shown in Fig. \ref{fig:Mott}.
Finally the metal-insulator transition occurs to the Mott insulating state.
The obtained phase boundary is consistent with the critical point 
$(U/D, V/D)\approx (2,0)$ for the two-orbital Hubbard model 
with the large difference of the bandwidths.\cite{Inaba}

Our results are in clear contrast to those for the conventional periodic
Anderson model,
where at half filling and particle-hole symmetry,
no Mott transition occurs even in the large $U_1$ and $U_2$ case, but one
always finds a Kondo insulator.
Since this model is obtained in the limit Hund's exchange $J=0$ 
in our extended periodic Anderson model, 
we can therefore conclude that the competition between metallic 
and the insulating states originates from the Hund coupling.
This observation may be relevant for the heavy fermion behavior 
observed in transition metal oxides
such as $\rm Ca_{2-x}Sr_xRuO_4$ ($0.2<x<0.5$) and 
$\rm La_{n+1}Ni_nO_{3n+1}$,
which is now under investigation.

\section{Effect of hole doping}\label{sec4}

\begin{figure}[htb]
\begin{center}
\includegraphics[width=7cm,clip]{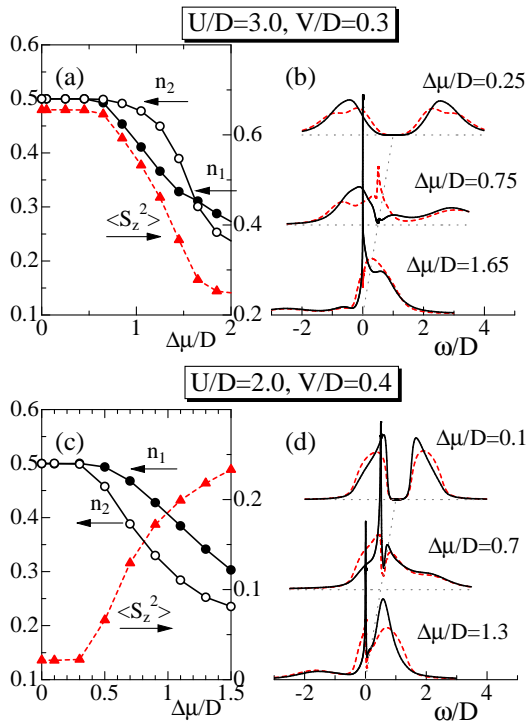}
\end{center}
\vskip -4mm
\caption{(color online)
Occupancy $\langle n_\alpha \rangle$
and squared moment $\langle S_z^2 \rangle$ 
as a function of the chemical potential (density of states)
for $V/D=0.2$ (a) [(b)] and $V/D=0.4$ (c) [(d)].
}
\label{fig:doping}
\end{figure}

In this section, we discuss the effect of hole doping 
on the insulating states to clarify how the
heavy fermion behavior emerges.
To this end, we introduce the parameter $\Delta\mu =\mu_0-\mu$.
In Fig.\ \ref{fig:doping}, we show occupancy, local moment
and density of states.
For $(U/D, V/D)=(3.0, 0.3)$, 
the system is in the Mott insulating state at half filling $\Delta\mu=0$,
and the increase of $\Delta\mu$ has little effect on 
the nature of the phase for $0.0< \Delta\mu/D <0.5$, 
as shown in Fig. \ref{fig:doping} (a).
Further increase of the chemical potential eventually drives the system into 
a metallic state,
where the local moment and the total occupancy $n_{tot}(=n_1+n_2)$ both decrease.
Note that most of the holes are doped in the $\alpha=1$ band while
only a small amount of holes populates the other 
when $0.5 < \Delta\mu/D < 1.0$. 
This implies that the character of original orbitals still survives
in the system with hybridization.
This orbital-selective behavior 
is also found clearly in the density of states, 
as shown in Fig. \ref{fig:doping} (b).
When $\Delta\mu/D>1.3$, the behavior changes: 
the number of electrons for the $\alpha=2$ band deviates
from the half filling, as shown in Fig. \ref{fig:doping} (a).
Furthermore, we find that a sharp peak appears in the density of 
states characteristic of the doped Kondo insulator. 
This suggests that a crossover between different metallic states
occurs around $\Delta\mu/D \approx 1.3$.

On the other hand, when holes are doped into the Kondo insulating state
at $(U/D, V/D)=(2.0, 0.4)$,
monotonic behavior appears in the quantities in Fig. \ref{fig:doping} (c). 
In this case, a conventional heavy fermion state,
studied in detail in the conventional periodic Anderson model, 
is realized, as shown in Fig. \ref{fig:doping} (d).

\begin{figure}[htb]
\begin{center}
\includegraphics[width=7cm,clip]{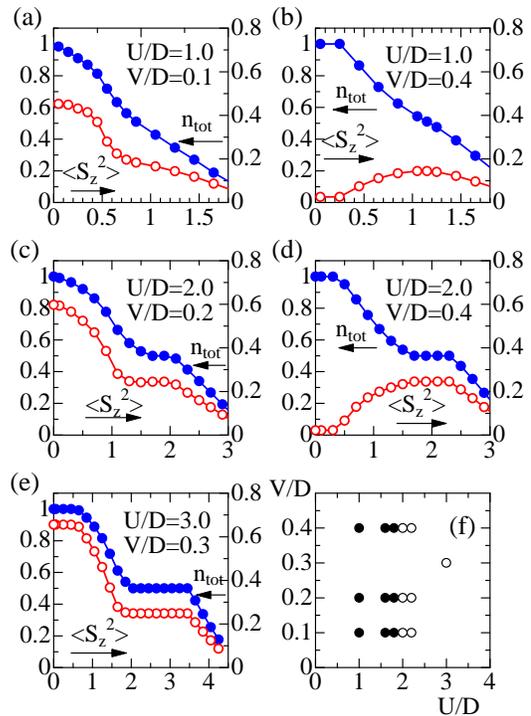}
\end{center}
\vskip -4mm
\caption{(color online)
(a)-(e) The total number of electrons 
$n_{tot}$ and local moment $\langle S_z^2 \rangle$ as a function 
of $\Delta \mu/D$ with fixed parameters at $T/D = 6.5\times 10^{-4}$. 
(f) Finite temperature phase diagram for the quarter filling
at $T/D = 6.5\times 10^{-4}$.
Solid (open) circles denote the metallic (Mott insulating) state.
}
\label{fig:quarter}
\end{figure}

Before closing this section, we briefly mention the Mott transitions 
at quarter filling, which should be induced by 
the intraorbital and the interorbital interactions.
The results obtained for several choices of parameters
are shown in Fig. \ref{fig:quarter}.
It is found that the ground state at quarter filling $n_{tot}=0.5$ 
strongly depends on the strength of the Coulomb interaction $U$.
When $U$ is small, 
no remarkable features are visible in the curve of the total occupancy,
as shown in Figs. \ref{fig:quarter} (a) and (b).
Therefore, we infer that the system is a metal state in the case.
In the large $U$ case, on the other hand, 
a plateau appears in the curve at quarter filling, 
as shown in Figs. \ref{fig:quarter} (c), (d) and (e).
In addition, 
one observes a local moment $\langle S_z^2 \rangle=1/4$ in this region,
which suggests the existence of a Mott insulating state.
Our results are similar to those for 
a two-orbital system without hybridization,\cite{Ono4,Inaba4}
which is a Mott insulator at quarter filling for large enough interaction parameters.
We can therefore conclude, that the hybridization has little effects on 
the phase diagram for the quarter filled system, 
as shown in Fig. \ref{fig:quarter} (f),
in contrast to that for the half-filled system (Fig. \ref{fig:phase30}).

\section{Summary}\label{sec5}
We have investigated the extended periodic Anderson model
within the dynamical mean-field theory with the numerical
renormalization group as a solver for the effective quantum impurity problem. 
We have discussed the nature and physics of phase transitions between
the metallic, Kondo insulating and Mott insulating states.
It has been clarified that the metallic and Mott insulating states are
stabilized by the Hund exchange coupling between the localized and itinerant
bands, which may be relevant to real materials, {\it e.g.}\ transition metal oxides
that show heavy-fermion behavior under certain conditions.

Although we have restricted our discussion to the paramagnetic state 
in the paper, it is an interesting and instructive problem 
to discuss possible instabilities toward ordered states such as 
magnetic order, superconductivity, etc.
In particular, an antiferromagnetic state induced by 
magnetic field\cite{Ohashi,Milat} can be of considerable interest in connection 
with various heavy fermion compounds.
These problems are currently under investigation.

\section*{Acknowledgments}
We would like to thank K. Inaba, S. Nakatsuji, M. Sigrist and Y. Maeno 
for useful discussions.
This work was partly supported by a Grant-in-Aid from the Ministry 
of Education, Science, Sports and Culture of Japan (NK, AK)
and the German Sciende Foundation (DFG) through the
collaborative research grant SFB 602 (TP, AK) and 
the project PR 298/10-1 (RP).  
AK would
in particular like to thank the SFB 602 for its support and hospitality during
his stays at the University of G\"ottingen.
Part of the computations were done at the Supercomputer Center at the 
Institute for Solid State Physics, University of Tokyo
and Yukawa Institute Computer Facility.

%

\end{document}